\documentclass[10pt,preprint]{aastex}
\usepackage{lscape}

\accepted{}
\journalid{}{}
\articleid{}{}

\shortauthors{Mohanty et al..}
\shorttitle{Spectroscopy of 2M0535A: Cool Spots vs. Opacity Uncertainties}

\begin{document}

\defcitealias{stassun06}{SMV06}
\defcitealias{stassun07}{SMV07}
\defcitealias{gomez09}{G09}
\defcitealias{chabrier07}{CGB07}
\defcitealias{mohanty09}{MSM09}
\defcitealias{mohanty04}{M04}
\defcitealias{reiners07}{R07}
\defcitealias{reiners05}{R05}

\def\lam{$\lambda$}
\def\tross{${\tau}_{R}$ }
\def\hal{{{\rm H}\alpha} }
\def\rc{R_C}
\def\ic{I_C}
\def\rcm{R_{Cm}}
\def\icm{I_{Cm}}
\def\jm{J_{m}}
\def\rco{R_{Co}}
\def\ico{I_{Co}}
\def\jo{J_{o}}
\def\av{A_V}
\def\ar{A_{R_C}}
\def\ai{A_{I_C}}
\def\exj{A_{J}}
\def\kr{k_{R_C}}
\def\ki{k_{I_C}}
\def\rad{{\mathcal{R}}_{\ast}}
\def\dist{\mathcal{D}}
\def\mass{{\mathcal{M}}_{\ast}}
\def\lum{{\mathcal{L}}_{bol}}
\def\logten{log_{10}}

\def\li{\ion{Li}{1}\ }
\def\na{\ion{Na}{1}\ }
\def\pot{\ion{K}{1}\ }
\def\oxy{\ion{O}{1}\ }
\def\hel{\ion{He}{1}\ }
\def\cal{\ion{Ca}{2}\ }
\def\nit{\ion{N}{2}\ }
\def\sul{\ion{S}{2}\ }

\def\ualph{${\mu}_{\alpha}$ }
\def\udelt{${\mu}_{\delta}$ }
\def\kms{km s$^{-1}$\ }
\def\kmsp{km s$^{-1}$ pix$^{-1}$\ }
\def\cms{cm s$^{-1}$\ }
\def\cmc{cm$^{-3}$\ }
\def\cmss{cm$^{2}$ s$^{-1}$\ }
\def\cmcs{cm$^{3}$ s$^{-1}$\ }

\def\mdot{$\dot{M}$}
\def\msun{M$_\odot$}
\def\rsun{R$_\odot$}
\def\lsun{L$_\odot$}
\def\mj{M$_{Jup}$ }

\def\teff{$T_{\rm eff}$}
\def\logg{$\log g$}
\def\tefflbol{T$_{e\! f\! f}$-L$_{\it bol}$~}
\def\gv{{\it g}~}
\def\vsini{{\it v}~sin{\it i}~}
\def\vrad{v$_{\it rad}$~}
\def\lbol{L$_{\it bol}$~}
\def\mbol{{\rm M}_{\it bol}}
\def\lhal{L_{H\alpha}}
\def\fhal{F_{H\alpha}}
\def\fcal{{\mathcal{F}}_{CaII}}
\def\fcont{{\mathcal{F}}_{cont}}
\def\fbol{F_{\it bol}}
\def\lx{L_{X} }
\def\eqwhal{EW_{H\alpha}}
\def\eqwcal{EW_{CaII}}
\def\alom{{\alpha}{\Omega} }
\def\ross{R_{0} }
\def\cots{{\tau}_{c} }
\def\fchal{{\mathcal{F}}_{c\hal} }
\def\h2o{H$_2$O}
\def\vcal{r_{CaII} }
\def\fex{{\mathcal{F}}_{excess}}

\title{High Resolution Spectroscopy during Eclipse\\ of the Young Substellar Eclipsing Binary 2MASS~0535$-$0546.\\ I.\ Primary Spectrum: Cool Spots versus Opacity Uncertainties}

\author{Subhanjoy Mohanty\altaffilmark{1},  Keivan G.\ Stassun\altaffilmark{2,3},
Greg W.\ Doppmann\altaffilmark{4}}

\altaffiltext{1}{Harvard-Smithsonian Center for Astrophysics, Cambridge, MA 02138, USA.  smohanty@cfa.harvard.edu}
\altaffiltext{2}{Department of Physics \& Astronomy, Vanderbilt University, Nashville, TN 37235, USA.  keivan.stassun@vanderbilt.edu}
\altaffiltext{3}{Department of Physics, Fisk University, Nashville, TN 37208, USA.}
\altaffiltext{4}{NOAO, 950 North Cherry Avenue, Tucson, AZ 85719, USA.  gdoppmann@noao.edu}

\begin{abstract} 
We present high-resolution Keck optical spectra of the very young substellar
eclipsing binary 2MASS J05352184$-$0546085, obtained during eclipse of the lower-mass 
(secondary) brown dwarf.
The observations yield the spectrum of the higher-mass (primary) brown dwarf alone, with negligible
($\sim$1.6\%) contamination by the secondary.  We perform a simultaneous
fine-analysis of the TiO-$\epsilon$ band and the red lobe of the \ion{K}{1}
doublet, using state-of-the-art PHOENIX {\sc dusty} and {\sc cond} synthetic spectra.
Comparing the effective temperature and surface gravity derived from these fits to the
{\it empirically} determined surface gravity of the primary (\logg\ = 3.5)
then allows us to test the model spectra as well as probe the prevailing
photospheric conditions.  We find that: {\it (1)} fits to TiO-$\epsilon$
alone imply \teff\ = 2500$\pm$50K; {\it (2)} at this \teff, fits to \ion{K}{1}
imply \logg\ = 3.0, 0.5 dex lower than the true value; and {\it (3)} at
the true \logg, \ion{K}{1} fits yield \teff\ = 2650$\pm$50K, $\sim$150K higher
than from TiO-$\epsilon$ alone.  On the one hand, these are the trends
expected in the presence of cool spots covering a large fraction of the primary's surface (as theorized previously to
explain the observed \teff\ reversal between the primary and secondary).
Specifically, our results can be reproduced by an
unspotted stellar photosphere with \teff\ = 2700K and (empirical) \logg\ =
3.5, coupled with axisymmetric cool spots that are 15\% cooler (2300K),
have an effective \logg\ = 3.0 (0.5 dex lower than photospheric), and
cover 70\% of the surface.  
%This appears an unusually large spot covering
%fraction, but is in line with the theoretically required coverage to cause
%the temperature reversal.  
On the other hand, the trends in our analysis can also be reproduced by
%may also arise from 
model opacity errors: there are lacks in the
synthetic TiO-$\epsilon$ opacities, at least for higher-gravity field dwarfs. 
%which, if confirmed and applicable to low-gravity young objects, would yield 
%a \teff spuriously low by $\sim$150--200K from TiO-$\epsilon$ alone, and an
%attendant underestimation of log[g] by $\sim$0.5 dex when coupled with \ion{K}{1}.
%Our fine analysis of the primary alone cannot distinguish between such
%opacity uncertainties and cool spots.  
Stringently discriminating between the
two possibilities requires combining the present results with an equivalent analysis of
the secondary (predicted to be relatively unspotted compared to the primary).

\end{abstract}


\section{Introduction\label{intro}}
2MASS J05352184$-$0546085 (henceforth 2M0535), a very young system located in
the Orion Nebula Cluster (ONC), has recently been identified by 
\citet[][hereafter SMV06]{stassun06} as the first known substellar eclipsing binary
(EB).  EBs allow exquisitely precise direct determinations of the component
masses and radii, and thus the surface gravities (\logg), 
as well as the ratio of their luminosities (or equivalently, ratio of their 
effective temperatures, \teff).  As such, 2M0535 permits the first stringent tests
of both the theoretical evolutionary models and the synthetic spectra that
are widely employed to characterize the vast majority of brown dwarfs (for
which direct measurements of mass, radius, and surface gravity are not possible).
%and temperature

The parameters of the system found by \citetalias{stassun06} were refined with
more data by \citet[][hereafter SMV07]{stassun07} and still further by
\citet[][hereafter G09]{gomez09}.  
SMV07 found a spectral type of M6.5$\pm$0.5 for the primary (higher-mass) component,
suggesting \teff\ $\approx$ 2700K \citep{golimowski04}.
The latest analysis by
G09 confirms the initial findings that: {\it (1)} Both components of 2M0535
are moderate mass brown dwarfs ($M_1$ = 0.0572$\pm$0.0033 \msun, $M_2$ =
0.0366$\pm$0.0022 \msun); {\it (2)} their radii ($R_1$ = 0.690$\pm$0.011
\rsun, $R_2$ = 0.540$\pm$0.009 \rsun) are consistent with the theoretical
prediction that young brown dwarfs of a given mass should be much larger
than their field counterparts\footnote{The 2M0535 component radii
are slightly---$\sim$10\%---underpredicted by some models \citepalias[see][]{stassun07};
the discrepancy becomes slightly stronger using the latest \citetalias{gomez09} radii cited
here compared to \citetalias{stassun07}'s numbers.}; and {\it (3)} the \teff\ ratio
of the components (\teff$_{,2}$/\teff$_{,1}$ = 1.050$\pm$0.004)
shows an unexpected reversal, with the primary being cooler than the lower-mass secondary. 
%The precise mass and radius measurements yield
%precise measurements of the surface gravities (\logg$_{,1}$ = 3.52$\pm$0.03,
%\logg$_{,2}$ = 3.54$\pm$0.03).

The reversal in temperatures is not predicted by any set of theoretical
evolutionary tracks.  To explain it, \citet[][hereafter CGB07]{chabrier07}
proposed that strong magnetic fields on the primary suppress its interior
convection and also produce cool surface spots; neither effect is included
in the standard evolutionary models, and both would act to depress its \teff. 
%compared to the standard prediction.  
\citet[][hereafter R07]{reiners07} 
subsequently found that, compared to the secondary, the primary is a
relatively fast rotator with strong chromospheric $\hal$ emission, which
supports the presence of strong magnetic fields in the latter.  \citetalias{gomez09} then
showed that the observed small-amplitude residual (non-eclipse) variations
in the 2M0535 lightcurve, modulated at the rotational periods of the
primary and secondary, can be well reproduced by cool spots asymmetrically
covering a small fraction ($\lesssim$ 10\%) of both components' surfaces.
While such small spots cannot explain the temperature reversal, \citetalias{gomez09}'s
analysis does indicate that spots are at least present.  Moreover, they
cannot rule out the very large ($\sim$ 50\% areal coverage) spots on the
primary required to explain the temperature reversal, as long as these are
arranged {\it symmetrically} about the rotation axis (e.g., polar spots,
equatorial bands, or `leopard spots').

Additionally, through analysis of the optical to mid-IR spectral
energy distribution (SED), \citet[][hereafter MSM09]{mohanty09} showed
that: ongoing accretion is highly unlikely in the 2M0535 system, lending
credence to the \citetalias{reiners07} conclusion that the $\hal$ emission in the primary
is chromospheric; and the system SED is consistent with effective
temperatures of [\teff$_{,1}$, \teff$_{,2}$] $\sim$ [2700, 2900]K, in
agreement with the mid-M spectral types of the components rather than with
the much lower $\sim$ [2300, 2450]K proposed by \citetalias{chabrier07} within their theory
of \teff\ reversal via magnetic field effects.  Combining their results with
the others cited here, \citetalias{mohanty09} concluded that while magnetically-induced
spot/convection effects probably do play an important role in determining
the \teff\ of the 2M0535 primary, as advocated by \citetalias{chabrier07} 
\citep[and as indeed seems to be the case for active {\it field} dwarfs;][]{morales08},
the theory is as yet insufficiently developed quantitatively, and small
age variations between the components may play a significant role as well
\citep[as in fact appears to be the case in another young EB, albeit with 
stellar-mass components, Par 1802;][]{stassun08}.

Finally, subsequent to MSM09's work, MacDonald \& Mullan (2009) have proposed a theory wherein magnetic fields inhibit the onset of convection (though do not suppress it entirely) throughout the 2M0535 primary, instead of just in the upper-most super-adiabatic layers as in the theory of CGB07.  The theory appears to reproduce the observations of 2M0535AB without invoking non-coevality or very large surface spots on the primary; this competing scenario must be evaluated as well.  

In this paper, we present Keck HIRES observations of 2M0535 obtained during
eclipse so as to isolate the spectrum of the higher-mass, lower-\teff\
primary brown dwarf. Comparing the observed spectrum with 
state-of-the-art brown-dwarf atmosphere models, we test the ability of
these models to correctly reproduce the accurately known surface gravity
\citepalias[\logg\ = 3.52$\pm$0.03;][]{gomez09}, and in the process we directly
probe the prevailing photospheric conditions of the primary brown dwarf
in the 2M0535 system.

\section{Observations and Data Reduction\label{obs}}
We observed 2M0535 on the night of UT 2007 Oct 23 with the 
High Resolution Echelle Spectrometer (HIRES) on 
Keck-I\footnote{Time allocation through NOAO via the NSF's
Telescope System Instrumentation Program (TSIP).}.
We observed in the spectrograph's ``red" (HIRESr) configuration
with an echelle angle of $-0.403$ deg and a cross-disperser
angle of 1.703 deg. In this configuration, the two features
of primary interest in this paper, TiO $\lambda\lambda$8435--8455 and
\ion{K}{1} $\lambda$7700, fall on the ``green" chip, in echelle orders 42 and 46, respectively.
We used the OG530 order-blocking filter and the 1\farcs15$\times$7\farcs0
slit, and binned the chip during readout by 2 pixels in the dispersion direction.
The resulting resolving power is $R\approx 34\, 000$, with a 3.7-pixel 
($\sim 8.8$ km s$^{-1}$) FWHM resolution element. 

We obtained three consecutive integrations of 2M0535, each of 2400~s.
ThAr arc lamp calibration exposures were obtained before and after
the 2M0535 exposures, and sequences of bias and dome flat-field exposures
were obtained at the end of the night. The 2M0535 exposures were 
processed along with these calibrations using standard 
IRAF\footnote{IRAF is distributed by the National Optical Astronomy
Observatory, which is operated by the Association of Universities for
Research in Astronomy (AURA) under cooperative agreement with the National
Science Foundation.}
tasks and the MAKEE reduction package written for HIRES 
by T.~Barlow. The latter includes optimal extraction of the
orders as well as subtraction of the adjacent sky background.
The three exposures of 2M0535 were processed separately
and then median combined with cosmic-ray rejection into a single final spectrum.
The signal-to-noise (S/N) of the final spectrum is $\sim 15$ per resolution element.

Importantly, we intentionally chose the observations to
coincide exactly with the secondary eclipse, i.e.\ when the lower-mass,
smaller, higher-\teff\ secondary component was behind the primary
as seen from Earth. The first exposure started at UT 12:20~h, 
and the third exposure ended at UT 14:22~h, corresponding to orbital phases of 
0.0709 and 0.0794, respectively, during which time the secondary is almost completely 
blocked \citepalias[cf.\ Fig.\ 3 in][]{stassun07}. Integrated over the entire 2-hr 
observation, the total light contribution from the secondary was $\approx$1.6\%.
The light contribution from the secondary was calculated using the accurately
determined radius ratio, temperature ratio, and orbital parameters, including the
orbital inclination, from the light curve modeling performed in G09.
Thus the resulting spectrum is effectively that of the primary alone.

\section{Synthetic Spectra\label{models}}
We use the latest version of synthetic spectra for plane-parallel atmospheres generated using the PHOENIX
code, designated AMES-Cond (version 2.4) and AMES-Dusty (version 2.4) \citep{allard01}.  
These synthetic spectra have become broadly used in the literature on low-mass stars
and brown dwarfs especially at young ages. In addition, these model spectra are
incorporated into the commonly used stellar evolution models of \citet{baraffe98} as well
as in the CGB07 models discussed in Sec.~\ref{intro}. Thus one of our aims in selecting
the PHOENIX synthetic spectra here is to assess these commonly used models in the context
of brown dwarf evolution.

The PHOENIX code \citep{hauschildt99} is a general purpose stellar atmosphere model tool
that makes use of very complex atomic models and line blanketing by hundreds of millions of atomic and
molecular lines.
The PHOENIX models used here incorporate the most recent AMES line lists for both TiO 
\citep{langhoff97,schwenke98} and H$_2$O \citep{partridge97}.  A good treatment
of H$_2$O is essential for analyzing optical spectra, even though H$_2$O
opacity dominates only in the infrared (TiO opacity is more important
in the optical).  This is because the overall H$_2$O opacity is larger,
and its lines occur closer to the peak of the SED than those of TiO, at
the low \teff\ in M spectral types.  Consequently, changes in the H$_2$O opacity
have a substantial effect on the atmospheric temperature structure and
thus on the emergent spectrum even in the optical.  A total of about 500
million molecular lines are currently included in the models; of these,
$\sim$207 million are lines of H$_2$O, and $\sim$172 million are of
TiO \citep{allard00,allard01}.  Here we use solar-metallicity models
([M/H]=0.0).  While the metallicity of 2M0535 is not explicitly known,
a large deviation from solar is not expected for a young object in a nearby star-forming region.  We discuss potential metallicity effects on our results in more detail in \S6.2.  

Dust formation is another potentially important effect in the low-temperature
atmospheres of M-type objects; grains affect the atmospheric structure as
well as the emergent flux.  Both models we examine treat grain formation
self-consistently, through chemical equilibrium calculations \citep[see][]{allard01}.
Under physical conditions where the chemical equations imply
{\it no} grain formation, the {\sc cond} and {\sc dusty} spectra are {\it identical};
in the models, this occurs for \teff\ $\gtrsim$ 2500K.  For the latter
temperatures, therefore, either set of synthetic spectra may be used.
The difference between the two is in their treatment of dust settling, once
grains have formed.  The two models represent the two limits of settling:
{\sc dusty} models treat the case where grains form and remain suspended in
the photosphere, while the {\sc cond} ones are applicable when dust has formed
but subsequently settled (``condensed'') out of the photosphere entirely.
Observations of field dwarfs indicate that dust settling becomes important
only in the L types \citep[e.g.,][]{schweitzer01}. For the mid-M spectral type of 2M0535,
dust {\it formation} may occur but the grains are likely to remain in the photosphere 
\citep[][\citetalias{mohanty04}]{jonestsuji97}.
%(Jones \& Tsuji, 1997; Mohanty et al. 2004).
We thus use {\sc dusty} models for \teff\ $<$ 2500K, and {\sc cond} models for \teff\
$\geq$ 2500K (where {\it no} dust forms)\footnote{As noted, {\sc dusty} and
{\sc cond} are identical at these \teff; at high-resolution, only {\sc cond} 
models are available for \teff\ $>$ 2500K, so we use them for
\teff\ $\geq$ 2500K after verifying that they are indeed nearly identical
to {\sc dusty} at the overlap \teff\ of 2500K.}.

\section{Methodology \label{analysis}}
We wish to determine the \teff\ and surface gravity
(\logg) of the higher-mass component of 2M0535 (the ``primary," hereafter 2M0535A) 
from comparisons to synthetic spectra.
As \citet[][hereafter M04]{mohanty04} have shown, two ideal
regions for this analysis are the TiO-$\epsilon$ bandheads at
$\lambda\lambda\lambda$8435,8445,8455, and the red-lobe of the \ion{K}{1}
doublet at $\lambda$7700 (the blue lobe falls in the gap between
echelle orders in the HIRES setting used).  In particular, the TiO
bandheads are very sensitive to \teff, but negligibly so to gravity,
while the \ion{K}{1} absorption is sensitive to both; using the two regions
in tandem therefore enables one to disentangle and individually determine these two parameters.

Comparing the data to models requires some modifications to both.
These are discussed in detail in \citetalias{mohanty04}; the salient points are as follows.
The models are rotationally broadened (using Gray's methodology (Gray 1992), with a standard limb-darkening parameter of 0.6) by 10 \kms to match the observed
$v\sin i$ of 2M0535A \citepalias{reiners07}, and further broadened by convolution with
a Gaussian profile to match the instrumental broadening (finite
resolution) of the data.  Since our data are not flux-calibrated,
comparison to the models also requires some form of scaling.  This is
accomplished by normalizing both the data and models by their average
flux over a narrow region of pseudo-continuum\footnote{`Pseudo-'
because there is no true continuum in such cool objects, only an
apparent continuum made up of millions of overlapping molecular
lines; in the interests of conciseness, we drop the `pseudo-'
appelation forthwith.} just outside the absorption lines of interest:
over $\lambda\lambda$[8402.5--8411.5] for the the TiO-$\epsilon$
region, and over $\lambda\lambda$[7707.5--7709.5] for the \ion{K}{1} region
(wavelengths in the laboratory rest-frame).  Recall that the data are
also flat-fielded, which removes the blaze-function but preserves the
innate shape of the stellar spectrum.  Our normalization procedure then
ensures that the data and models are only `anchored' over a narrow
wavelength range, but otherwise unconstrained, so the models need
to match not only the absorption bands, but also the shape and slope
of the continuum, to ensure a good fit.  This provides an additional
check on the veracity of the preferred fits.

For comparisons to the stellar spectrum, we first use the {\sc dusty} and
{\sc cond} models depending on the \teff\, being tested, as described in \S\ref{models}
({\sc cond} for $\geq$ 2500K); the results are
described in \S\ref{results}.  We then model the stellar spectrum as a combination
of a naked photosphere and cool spots, as discussed in \S\ref{disc:spots}.  For this
analysis, the photosphere and spot are represented by different
spectra, depending on the adopted temperature of each (e.g., for a
2700K photosphere and a 2300K cool spot, we use {\sc cond} for the former
and {\sc dusty} for the latter).  Both spectra are individually rotationally
and instrumental-Gaussian broadened, and then coadded in the ratio
of the adopted spot covering fraction (for a covering fration $f$,
the final spectrum is given by $f$ times the spot spectrum + $(1-f)$
times the photospheric spectrum).  The addition is performed {\it
prior} to the normalization described above, to preserve the ratio
of the spot to photospheric flux arising from their differences in
both temperature and covering fraction.

Finally, we note that the synthetic spectra were originally constructed at intervals of 100K and 0.5 dex in \teff\ and \logg\ respectively.  We have linearly interpolated between adjacent spectra (before normalization) to construct a finer final grid of models, with steps of 50K in \teff\ and 0.25 dex in \logg.  

\section{Results\label{results}}
We begin with a general presentation of the fitting results and then quantify the uncertainties in the best fits.

\begin{figure}[h]
\epsscale{0.9}
\plotone{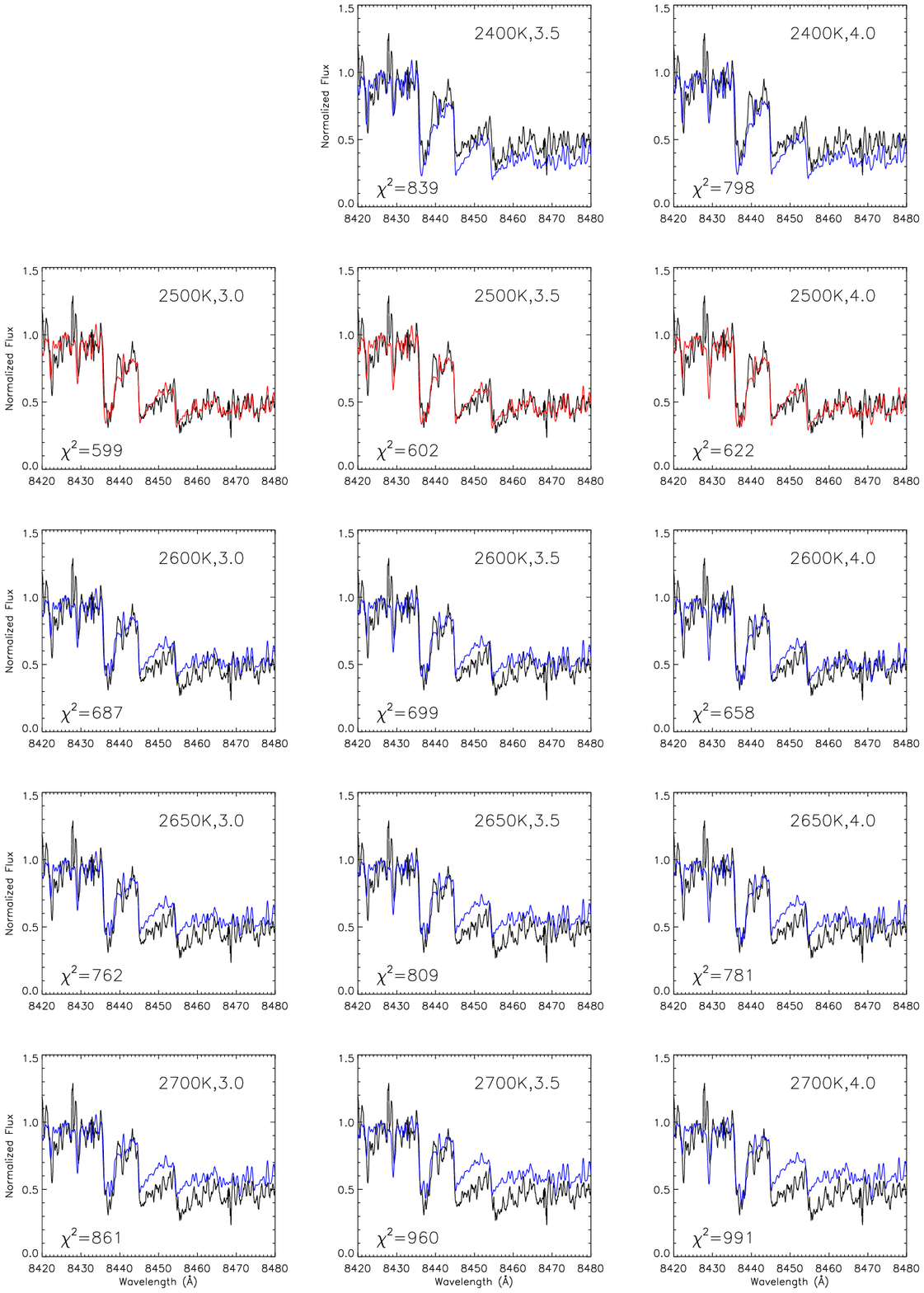}
\caption{\label{fig1} Observed TiO-$\epsilon$ region in 2M0535A
({\it black}) compared to Dusty ($<$ 2500K) and Cond ($\ge$ 2500K)
models.  Best-fit model (Cond 2500K) shown in {\it red}; all others,
which clearly diverge from the data by eye, shown in {\it blue}.
Note that the model fits are very insensitive to gravity over the 1
dex range plotted. See \S5. This figure is shown in color in the electronic
version only.}
\end{figure}

\noindent {\bf TiO-$\epsilon$}: Fig.~\ref{fig1} shows the comparison between data and
synthetic spectra in the TiO-$\epsilon$ region, which has three bandheads
at $\sim$ $\lambda\lambda\lambda$8435,8445,8455.  We plot models at \logg\
= 3.0, 3.5 and 4.0 (bracketing the empirically known value of 3.5), and
\teff\ = 2400--2700K in steps of 100K (2650K is also shown to facilitate a
comparison to \ion{K}{1}, as we will discuss shortly).  As expected, the
model TiO is hardly sensitive to gravity over the 1~dex range plotted, but
highly sensitive to temperature, with the bandheads at
$\lambda\lambda$8445,8455 rapidly strengthening with decreasing \teff.  We see that the 2500K model (in {\it red}) clearly
fits the data very well, while cooler and hotter models (in {\it blue})
just as clearly do not.  Given that even a 100K deviation from the best
fit is evident to the eye, our precision in \teff\ determination by eye is likely to be $\sim\pm$50K (in agreement with M04). From the TiO-$\epsilon$ fits,
therefore, we would infer \teff\ $\approx$ 2500K$\pm$50K.

\begin{figure}[h]
\epsscale{0.85}
\plotone{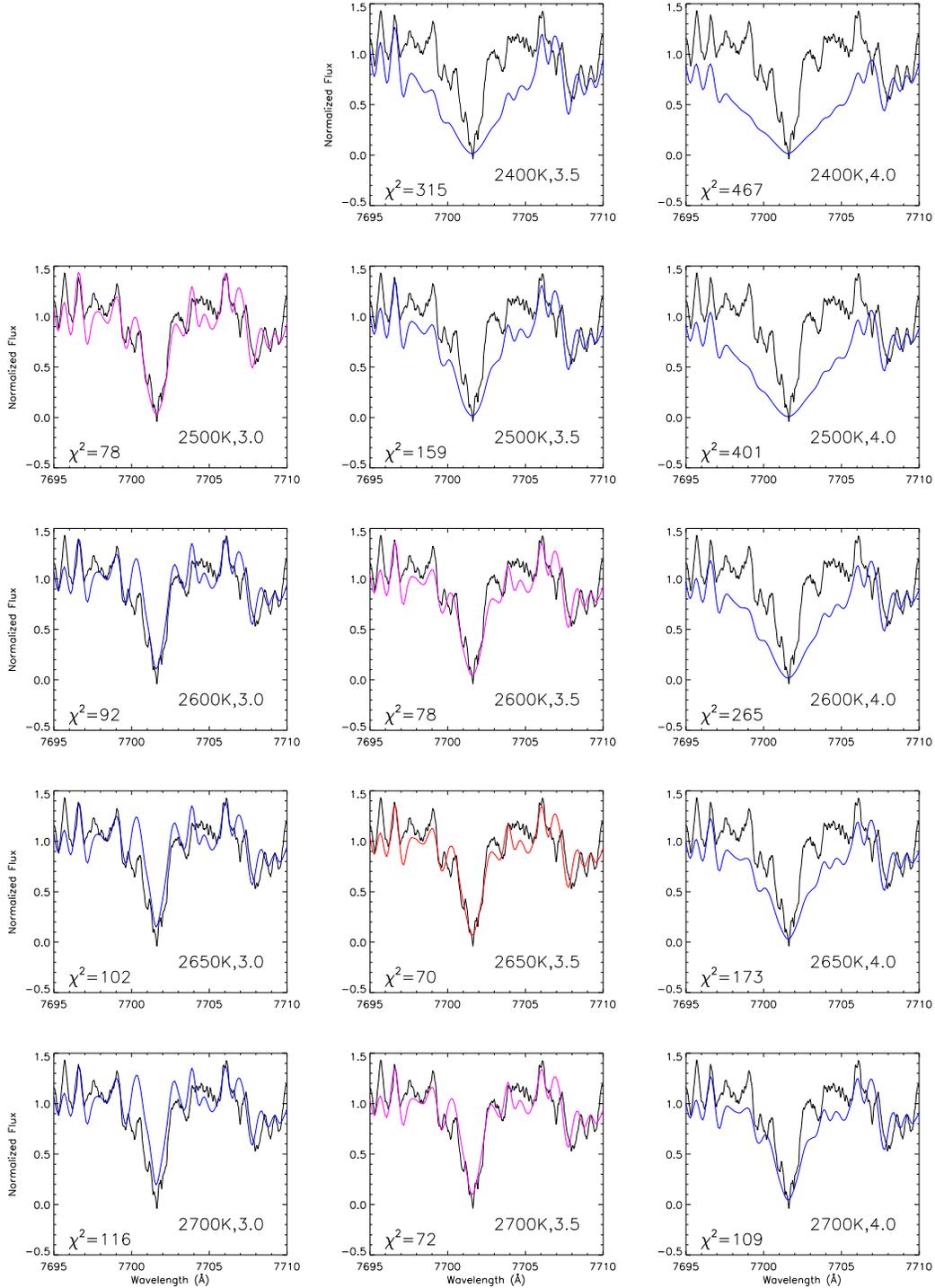}
\caption{\label{fig2} Observed red lobe of K I in 2M0535A ({\it
black}) compared to Dusty ($<$ 2500K) and Cond ($\ge$ 2500K) models.
Best-fit models (Cond [2500K, log $g$ = 3.0] and [2650K, log $g$ =
3.5]) shown in {\it red}; worse but still admissible fits by eye shown
in {\it purple}; and all others, which clearly diverge from the data,
shown in {\it blue}.  Note that the K I absorption is sensitive to
both \teff\, and gravity: a 150K increase in \teff\, compensates for
a 0.5dex rise in log $g$.  At \teff\, = 2500K, corresponding to the
best-fit to TiO-$\epsilon$ (Fig. 1), the K I implies log $g$ = 3.0,
while at the empirically determined log $g$ = 3.5, it implies \teff\,
= 2650K.  See \S5. This figure is shown in color in the electronic version only.}
\end{figure}

\noindent {\bf \ion{K}{1} and TiO-$\epsilon$}: Fig.~\ref{fig2} shows the comparison
between the observed and model \ion{K}{1} $\lambda$7700, over the same range of
\logg\ and \teff\ plotted in Fig.~\ref{fig1} for TiO-$\epsilon$.  We immediately
see, as pointed out by \citetalias{mohanty04}, that the synthetic \ion{K}{1} is sensitive to both
gravity and temperature, becoming rapidly broader and deeper with either
increasing gravity or decreasing \teff.  At the \teff\ = 2500K inferred from
TiO-$\epsilon$ above, the best fit to \ion{K}{1} is clearly obtained at \logg=3.0.  
The model at \logg\ = 3.5, for the same \teff, is obviously discrepant with the data; a conservative estimate of the error in this by-eye fit is $\sim$$\pm$0.25 dex \citepalias[in agreement with][]{mohanty04}.
Thus, simultaneous model fits to TiO-$\epsilon$ and \ion{K}{1} imply \teff\ 
$\approx$ 2500K and \logg\ $\approx$ 3.0.  This inferred gravity is lower
than the empirical value of 3.52$\pm$0.03 by $\sim$0.5 dex.

Fig.~\ref{fig2} also shows that, if we {\it impose} the empirical \logg\ of
3.5, a very good fit to \ion{K}{1} is obtained at \teff\ $\approx$ 2650K.
This is $\sim$150K higher than derived from the TiO-$\epsilon$ fits alone;
as Fig.~\ref{fig1} shows, the synthetic TiO-$\epsilon$ are incompatible with this
temperature.  We note that 2650K is consistent with \teff\ estimates for
field dwarfs of the same $\sim$M6.5 spectral type as the primary 
\citep[e.g.,][]{golimowski04,slesnick04}, 
while the 2500K indicated by TiO-$\epsilon$ is somewhat low in comparison.

To better quantify the uncertainties in the fitting results of Figs.\ 1 and 2, we show the chi-square results for the data-model comparisons in Fig.~3.  The TiO comparisons were carried out over the wavelength range [8420:8480]\AA~ (which includes all three bandheads; see Fig. 1), while the KI comparisons were over the range [7699:7706]\AA~ (corresponding to the entire line, till the pseudo-continuum is reached on either side of line-center; see Fig. 2).  These ranges correspond to 69 data points for KI, and 599 for TiO.  

The plot clearly shows that a degeneracy exists between \teff\ and \logg\ in the KI line, as discussed above.  There are two global minima over the range of temperatures and gravities examined, one at \teff\ $\approx$ 2650K and \logg\ $\approx$ 3.5, and another at \teff\ $\approx$ 2850K and \logg\ $\approx$ 4.0.  That is, a $\sim$200K decrease in temperature can compensate for a 0.5 dex decrease in gravity (as also found by M04).  Thus, at the empirical \logg\ of 2M0535, (\logg\ = 3.5), the best-fit \teff\ from KI is 2650K, in agreement with our by-eye estimate above. The corresponding 1-$\sigma$ uncertainty is 30K. The TiO lines strongly indicate \teff\ = 2500$\pm$10K, again as found above.  At this \teff, the \ion{K}{1} line is marginally well fit at the 3-$\sigma$ contour level with \logg\ = 3.0$\pm$0.05.  Thus, the detailed chi-square comparisons are in excellent agreement with our fits-by-eye results for \teff\ and \logg.

To summarize: {\bf (1)} fits to TiO-$\epsilon$ yield \teff\ = 2500$\pm$10K;
{\bf (2)} adopting this \teff, fits to \ion{K}{1} $\lambda$7700 imply \logg\
= 3.0$\pm$0.05, {\it which is $\sim$0.5 dex lower than the known gravity of
the primary}; and {\bf (3)} adopting the known value of \logg\ = 3.5 instead,
fits to \ion{K}{1} imply \teff\ = 2650$\pm$30K, {\it which is $\sim$150K
higher than, and incompatible with, the value from TiO-$\epsilon$ alone}. 

We note here that, while the best-fit chi-square values are the same as obtained by eye, the formal uncertainties cited above for the chi-square analysis, obtained via interpolation over the model grid, are significantly smaller than the model grid spacing (50K in \teff\ and 0.25 dex in \logg).  We therefore adopt the grid spacing of 50K and 0.25 dex as a conservative estimate of our uncertainties for the rest of the paper (corresponding to the same errors assumed for the fits by eye).  

The discrepancies embodied in the above results may be due to lacks in
the synthetic spectra, or an indication of real photospheric conditions.
We discuss each in turn below.

\section{Discussion: 
Possible Interpretations of the Discrepancies in $T_{\rm eff}$
and $\log g$ \label{discussion}}
\subsection{Model Opacity Uncertainties\label{disc:opacity}}
\citet[][hereafter R05]{reiners05} has compared the synthetic TiO spectra to
observations of a sample of early to mid-M field dwarfs, whose \teff\
and surface gravities are well-constrained via interferometric radius
measurements.  He shows that the model TiO-$\epsilon$ bands systematically
underestimate the temperatures of these objects.  Assuming that uncertainties in the $\epsilon$-band model oscillator 
strengths---$f_{\rm el}$($\epsilon$)---are to blame, 
\citetalias{reiners05} estimates that $f_{\rm el}$($\epsilon$) 
70\% higher than adopted in the models would remove
the discrepancy in the field dwarfs.  Moreover, problems with 
$f_{\rm el}$($\epsilon$) should produce an analogous effect in young 
low-gravity brown dwarfs as well.  \citetalias{reiners05} predicts that a 70\% underestimation 
of $f_{\rm el}$($\epsilon$) would yield a 150--200K underestimation of 
\teff\ in such young dwarfs, and an attendant \logg\ (derived by imposing this \teff\ 
on the gravity-sensitive alkali lines, as in our analysis above) too low by $\sim$0.3 dex.
These results are in qualitative and quantitative agreement with the results presented above for 2M0535.

The above prediction for young cool dwarfs is predicated on the assumption
of problematic oscillator strengths.  This is by no means proved,
however; it may be that the fault lies in the adopted model equation
of state instead (A.~Reiners and P.~Hauschildt, private comm., 2009),
perhaps related to uncertainties in dust formation.  In the latter case,
it is not evident that the field dwarf discrepancies would necessarily be
replicated in young objects.  The steps required to resolve this issue are
discussed further in \S\ref{summary}.  For now, we simply point out that the direction
and magnitude of the $f_{\rm el}$($\epsilon$)-related \teff\ and \logg\
discrepancies predicted by R05 for young brown dwarfs {\it are} consistent
with those found in our analysis of the 2M0535 primary.  As such, lacks
in the synthetic spectra remain a viable explanation for our results.

\subsection{Metallicity Effects}
Alternatively, we have assumed a solar metallicity for the 2M0535 system (i.e., used synthetic spectra with log[M/H] $\equiv$ log[(M/H)/(M$_\odot$/H$_\odot$)] = 0.0); can non-solar abundances resolve the \teff\ and \logg\ discrepancies we find?  We do {\it not} think so, for the following reason.

Metallicity variations affect the spectra as follows.  Higher metallicity reduces the number of Hydrogen particles (which are the main source of collisional broadening) relative to metals; it also implies a decrease in pressure at a given optical depth (because of higher opacity).  Both effects tend to yield a narrower alkali line at higher metallicity, just as {\it decreasing} gravity does at solar metallicity (Schweitzer et al. 1996; Basri et al. 2000; Mohanty et al. 2004).  
This could lead to an underestimated \logg\ from \ion{K}{1}.
Simultaneously, higher metallicity leads to an increase in the relative abundance of Titanium and Oxygen, and also causes a decrease in temperature at a given optical depth (again due to higher opacity).  Both effects lead to a strengthening of the TiO bands at higher metallicity, just as {\it decreasing} \teff\ does at solar metallicity (Leggett 1992; Mohanty et al. 2004).  
In summary, a higher metallicity mimics lower gravity and lower \teff.  Thus, if we have underestimated the metallicity, we will also erroneously underestimate \teff\ and \logg\ to compensate (i.e., to match the observed line profiles).

These effects may be potentially invoked to explain our results in the following way.  If metallicity in the 2M0535 system is higher than solar, then accounting for this will produce a \teff\ (from the TiO bands) that is somewhat higher than we currently find assuming solar abundances.  Simultaneously, using the putative, higher-than-solar abundance would also lead to a higher \logg\ (at any chosen \teff), from the KI line analysis, than we find at present.  If the metallicity were sufficiently higher than solar, then these trends could potentially lead to an agreement in \teff\ between the TiO and KI regions at the correct (empirically known) gravity, resolving the discrepancies in our present results.  

However, it is the {\it magnitude} of the metallicity change required that is the stumbling block.  On the one hand, Padgett (1996) has analyzed a number of nearby star-forming regions (Taurus-Auriga, Ophiuchus, Chameleon and Orion, the latter being the region of which 2M0535 is a member) and found a solar abundance to within $\pm$0.1 dex (i.e., $\pm$25\%) in all of them; within a given region, the variation is also at most 0.1 dex.  On the other hand, Schweitzer et al. (1996) have shown that, for a fixed \teff\ ($\sim$ 2700K, i.e., around the \teff\ regime of interest here), a 0.5 dex increase (decrease) in log[M/H] mimics a 0.5 dex decrease (increase) in \logg\ in the synthetic alkali line profiles.  Similarly, from our synthetic spectra, for a fixed \logg\, we find that a 0.5 dex increase (decrease) in metallicity mimics a $\sim$100K decrease (increase) in \teff\ in the synthetic $\epsilon$-band TiO \footnote{Only low-resolution, \logg\ = 5.0 synthetic spectra are currently available for non-solar metallicities in the {\sc cond} and {\sc dusty} models.  However, the \teff\ versus log[M/H] changes are clear even at low-resolution, and the insensitivity of the TiO $\epsilon$-band to \logg\ indicates that these results should be reasonably applicable at the lower gravity of our sources as well.}.  

Thus, the 0.1 dex maximum observed variation in metallicity would lead to non-significant changes in our inferred \teff\ and \logg : $\sim$20K and 0.1 dex respectively.  These deviations are less than or comparable to the error bars on our derived values, and more importantly, completely insufficient to explain the discrepancies in temperature and gravity we find between the TiO and KI regions.  We therefore posit that metallicity variations are {\it not} likely to explain our results, as an improbably large $[M/H] \gtrsim 0.5$ dex for 2M0535 would be required.

\subsection{Cool Spots\label{disc:spots}}
Finally, as discussed in \S\ref{intro}, \citetalias{chabrier07} 
predict a significant presence
of cool spots on the primary, to explain its \teff\ reversal compared to
the secondary.  Assuming an admittedly extreme spot temperature of 0~K,
they require a spot covering fraction of $\sim$50\% (combined with severe
magnetically-induced suppression of interior convection) to replicate the
observations.  More recently, adopting more realistic spots 10\% cooler than
the bare photosphere in their photometric lightcurve analysis, 
\citetalias{gomez09} find that a spot coverage of $\sim$65\% can reproduce
the observed \teff\ suppression of the primary (though, as mentioned in \S\ref{intro},
these spots must be distributed axisymmetrically, to remain consistent with
the relatively small photometric {\it variations} \citetalias{gomez09} observe; the latter
require only $\lesssim$10\% coverage by {\it non}-axisymmetric spots).

In light of this, we investigate the effects of cool spots on our spectra.
A priori, the following trends are expected.  First, since the spots are
by definition cooler than the unspotted photosphere, the TiO bandheads
arising inside them will be deeper (relative to the continuum) than
those from the surrounding photosphere; the resultant average TiO in a
spatially unresolved spectrum will then imply a temperature intermediate
between that of the spotted and unspotted surfaces.  This trend will not
however be monotonic with decreasing spot temperature, since the 
{\it continuum flux} from spots also falls with the spot temperature: the spot TiO
contributes less to the average TiO as the spot gets cooler.  Thus, for a
fixed unspotted photospheric \teff\ and spot coverage, the average TiO in
the combined spotted$+$unspotted spectrum first deepens (relative to the
TiO from an unspotted surface) with decreasing spot temperature, and then
reverses as the spots become still cooler, to become shallower again (i.e.,
approaches again the TiO from an unspotted surface).  The spot temperature at which this reversal occurs, and thus the maximum depth of the TiO bandheads in the presence of spots, is determined by both the unspotted photospheric \teff\ and the spot coverage assumed.  As an extreme example,
0~K spots contribute no flux at all, and hence, for any coverage $<$100\%,
will have {\it no} effect on the {\it shape} of the TiO, which arises in
this case only from the unspotted surface (even though the absolute flux
in the continuum and bandheads will be lower than in the absence of spots, since the unspotted surface now covers only a fraction of the total stellar surface).  The trend in TiO with changing spot temperature and coverage is illustrated in the top panel of Fig. 4.  

Second, inside a spot, magnetic fields provide partial support against
the external photospheric gas pressure; consequently, the gas pressure
alone within a spot is lower than in the surrounding unspotted surface
\citep[e.g.,][]{amado99,amado00}.  This exactly mimics the reduction of gas
pressure caused by lower surface gravity, and causes the gravity- (more
accurately, gas-pressure-) sensitive alkali absorption lines 
(e.g.\ \ion{K}{1}) in a cool spot
to be narrower than outside, thereby implying a lower {\it effective}
gravity within the spot.  The averaged alkali lines in the combined
spotted$+$unspotted spectrum will then imply a gravity intermediate between
the true photospheric value and the effective one in the spot.  Conversely,
spots are also cooler than the external photosphere; this tends to make
the alkali lines, which are temperature-sensitive as well, {\it broader}
within a spot, mitigating the narrowing caused by the magnetic pressure
effects and reducing the apparent gravity offset.  Finally, both effects
are limited by the decreasing flux from a spot with lower temperature,
exactly as discussed above for TiO; for a given unspotted photospheric
\teff\ and spot coverage, the {\it shape} of an alkali absorption line is
negligibly changed, relative to that from an unspotted surface, once the
spot temperature falls below a certain threshold.  The trend in KI with changing spot temperature and coverage is illustrated in the bottom panel of Fig. 4.  

These trends have the following consequences.
As we have shown, the maximum depth of the TiO bandheads in the presence of spots is determined by both the unspotted photospheric \teff\ and the spot coverage adopted.  Thus, for an observed TiO band depth, specifying the unspotted photospheric \teff\ sets a lower limit to the spot areal coverage: if the coverage is below this value, the average TiO band depth reverses before it can match the observed depth regardless of how cold
the spots become.  For a coverage higher than this threshold and fixed
unspotted photospheric \teff, there is a degeneracy between spot
temperature and coverage: a larger coverage allows cooler spots.  At the
same time, if we decrease the spot temperature we must also decrease 
the spot's effective gravity in order to match the observed alkali absorption lines, 
given the competing effects of gravity and temperature in these lines.

These trends imply that, with a priori knowledge of the unspotted
\teff, and the true stellar gravity as well as the effective gravity within a spot, one can solve for the
spot temperature as well as covering fraction via simultaneous analysis
of the TiO and alkali absorptions. We however only know the stellar gravity, and have no advance knowledge of either the unspotted photospheric \teff\ or the spots' effective gravity.  As such, we can make no claims to a unique solution, or to a full search
of the available parameter space.  Instead, our goal is a plausible
solution, based on constraints set by known properties of sunspots and
starspots in general as well as by the 2M0535 lightcurve analysis so far.
In particular, we assume that the spots are cooler than the surrounding
photosphere by at most $\sim$10--25\% 
\citep[e.g., \citetalias{gomez09};][]{linsky02},
and that the differential between their effective gravity and the higher 
photospheric value (\logg=3.5) is $\lesssim$ 0.5 dex \citep[e.g.,][]{amado99,amado00}.  The unspotted photospheric \teff\ is also kept a free parameter, but with a lower bound of 2500K set by the \teff\ inferred from the TiO fits in \S5 (the reason for this lower bound is that, if the unspotted photosphere \teff\ were $<$ 2500K, then combining it with spots that are by definition even cooler could never produce TiO bands that appear to be at 2500K when compared to unspotted models).  

Within these constraints, we construct star$+$spot models by assigning
synthetic spectra to the unspotted and spotted surfaces, as detailed in \S\ref{analysis}.
We find a viable solution for our spectroscopic data with the following
parameters: an unspotted stellar surface with \teff\ = 2700K and (empirically
determined) \logg\ = 3.5, combined with 70\% axisymmetric areal coverage
by spots with a temperature of 2300K and effective \logg\ = 3.0 (i.e.,
15\% cooler, and 0.5 dex lower apparent gravity, than the unspotted surface).

Fig. 5 demonstrates the viability of this spotted model.  In the left panels, we compare the data (in {\it black}) to the best-fit unspotted models from Figs. 1 and 2 (TiO: best-fit \teff\ of 2500K [in {\it red}], as well as 2550K [in {\it grey}] to show the deviation caused by our adopted 50K uncertainty; KI: best-fit \logg\ = 3.0 at the 2500K implied by TiO [in {\it blue}], as well as best-fit \teff\ = 2650K at the empirically determined \logg\ of 3.5 [in {\it red}]).  In the right panels, we compare the best-fit spot model described above (in {\it green}) to both the data and the best-fit unspotted models from the left panels.  We see that the spotted model very closely reproduces the data as well as the unspotted model fits (in particular, the spotted model is identical to the 2550K unspotted model in TiO, i.e., within 50K -- our adopted error -- of the best-fit 2500K unspotted TiO model; it is also nearly indistinguishable from the 2650K KI model at the empirical gravity of \logg\ = 3.5).   

Thus, the discrepancies in \teff\ and \logg\ implied by the unspotted model-fits to TiO-$\epsilon$ and \ion{K}{1} are resolved by this single star+spot model.  Consequently, cool spots are a viable explanation for our data.

\section{Summary and Conclusions\label{summary}}

We have shown that the TiO-$\epsilon$ and \ion{K}{1} absorption features in
the high-resolution optical spectrum of the higher-mass primary in 2M0535
(2M0535A) are consistent with a
\teff\ = 2700K, (empirical) \logg\ = 3.5 photosphere combined with cool spots
with a temperature of 2300K and effective gravity of \logg\ = 3.0
covering 70\% of the brown dwarf's surface.  This is in agreement with
the scenario outlined by \citetalias{chabrier07}, wherein the temperature reversal in
the primary relative to the secondary is caused by magnetic fields,
which induce both a reduction in the convective efficiency and high
cool spot coverage.  While the extreme spot covering fraction we
find is similar to that inferred by \citetalias{chabrier07} (50\% using unrealistic 0~K
spots) and \citetalias{gomez09} (65\% with more realistic spots 10\% cooler than the
photosphere), this very high fraction is nevertheless troubling:
in effect, it makes 2M0535A appear to be a ``very cool'' (2300K)
star covered by hot spots, rather than the reverse.  On the other hand, we note that Stauffer et al. (2003) argue that the anomalous colours of Pleiades K and M dwarfs result from axisymmnetrically distributed cool spots with a very large areal coverage, $\geq$ 50\%; Gullbring et al. (1998) have argued for similarly large axisymmetric spots, with covering fractions of $\sim$ 50--70\%, to account for the anomalous colors of even younger weak-lined T Tauri stars (WTTS).  There is also evidence for the presence of such axisymmetric large spots from Doppler imaging of WTTS, e.g., large polar spots in V410 Tau (Hatzes 1995) and HDE 283572 (Joncour et al. 1994).  Thus, such spots may be usual during the early evolution of these stars, when they are rapidly rotating and highly active, and the phenomenon may extend into the substellar regime as well.  

The other option, which we show our results are also consistent with,
is that the synthetic spectra are in error, leading to a discrepancy
between the \teff\ and \logg\ derived from simultaneous fits to
TiO-$\epsilon$ and \ion{K}{1} (which then leads us to postulate a surfeit
of cool spots).  \citetalias{reiners05} postulated such synthetic spectrum errors, also
seen in analyses of field dwarfs, to arise from problems with the model
TiO-$\epsilon$ oscillator strengths.  As we were submitting this paper,
it came to our attention that the model errors may lie in the adopted
equation of state instead, and that newer models rectifying this
are being prepared (A. Reiners \& P. Hauschildt, private comm., 2009).
Whether these models can resolve the discrepant values obtained
from TiO-$\epsilon$ and \ion{K}{1} in the case of the very young 2M0535A,
without having to resort to copious cool spots, remains to be seen.

Regardless of whether the new models fare better or not, setting rigorous constraints 
on both the models and the physical conditions on 2M0535A---i.e., determining whether 
the model fits (and thus the implied \teff\ and effective log $g$ for the spots) are truly valid and/or if 
very large cool spots exist on the primary---now requires independently carrying out 
exactly the same analysis for the lower-mass secondary (2M0535B).
%no single
%comparison between models and the 2M0535A data can lay the issue to
%rest, since the validity of the synthetic spectra will then still not
%have been tested independently.  
%The only way to truly test the models
%and concurrently probe conditions on 2M0535A is to carry out exactly
\citetalias{chabrier07}'s theory predicts that the secondary should be much less
spotted than 2M0535A:  thus, if the same discrepancies between
TiO-$\epsilon$ and \ion{K}{1} appear in the secondary as well, then errors
in the models will be clearly implied; if not, then the suggestion
that the primary has an extremely large spot covering fraction will
be bolstered.  We are currently undertaking observations of 2M0535B to carry out this test.

Finally, if it turns out from 2M0535B's analysis that the current
models {\it are} in error (i.e., if the same discrepancies between TiO and \ion{K}{1} 
appear in 2M0525B as well even though it is expected to be relatively unspotted), 
and thus there is no observational rationale for suggesting very large cool spots on 
the primary component 2M0535A, then the cause of the temperature reversal between the two binary
components once again becomes an unresolved issue.  In that case, it may be
that the theory proposed very recently by \citet{macdonald09}---wherein 
magnetic fields are again to blame, but by inhibiting
the onset of convection (though not completely) {\it throughout}
the star, instead of just in the uppermost super-adiabatic layers as
in the theory of \citetalias{chabrier07}---may be the correct one.  Again, we stress that
this can only be tested via comparison with an analogous analysis of 2M0535B.
Finally, complementary high-resolution spectroscopic observations in the near-infrared (NIR) can be very useful in determining the properties of cool brown dwarfs \citep[e.g.][]{zapatero06,mclean07,delburgo09}, as brown dwarfs are brightest in the NIR.  This would be particularly useful for a comparative analysis for the 2M0535 system, where there is a more favorable contrast ratio between the secondary and the primary in the NIR; it would also be helpful for assessing the importance of cool spots, whose influence is less marked in the NIR than in the optical.

This binary, the first eclipsing system in the substellar domain,
has already proved to be a rich testing ground for theories of brown
dwarf formation and evolution, and it promises to continue being a
Rosetta Stone in this regard.

\acknowledgments
S.M. wishes to thank G. Chabrier, I. Baraffe, A. Reiners and D.J. Mullan for stimulating discussions on the subject, and F. Allard
and P. Hauschildt for supplying the high-resolution synthetic
spectra.  K.G.S.\ acknowledges support from NSF grant AST-0607773.  
This material is based upon work supported
by AURA through the National Science Foundation under AURA Cooperative
Agreement AST-0132798 as amended.

%\clearpage

%\clearpage

\begin{figure}[h]
\epsscale{0.9}
\plotone{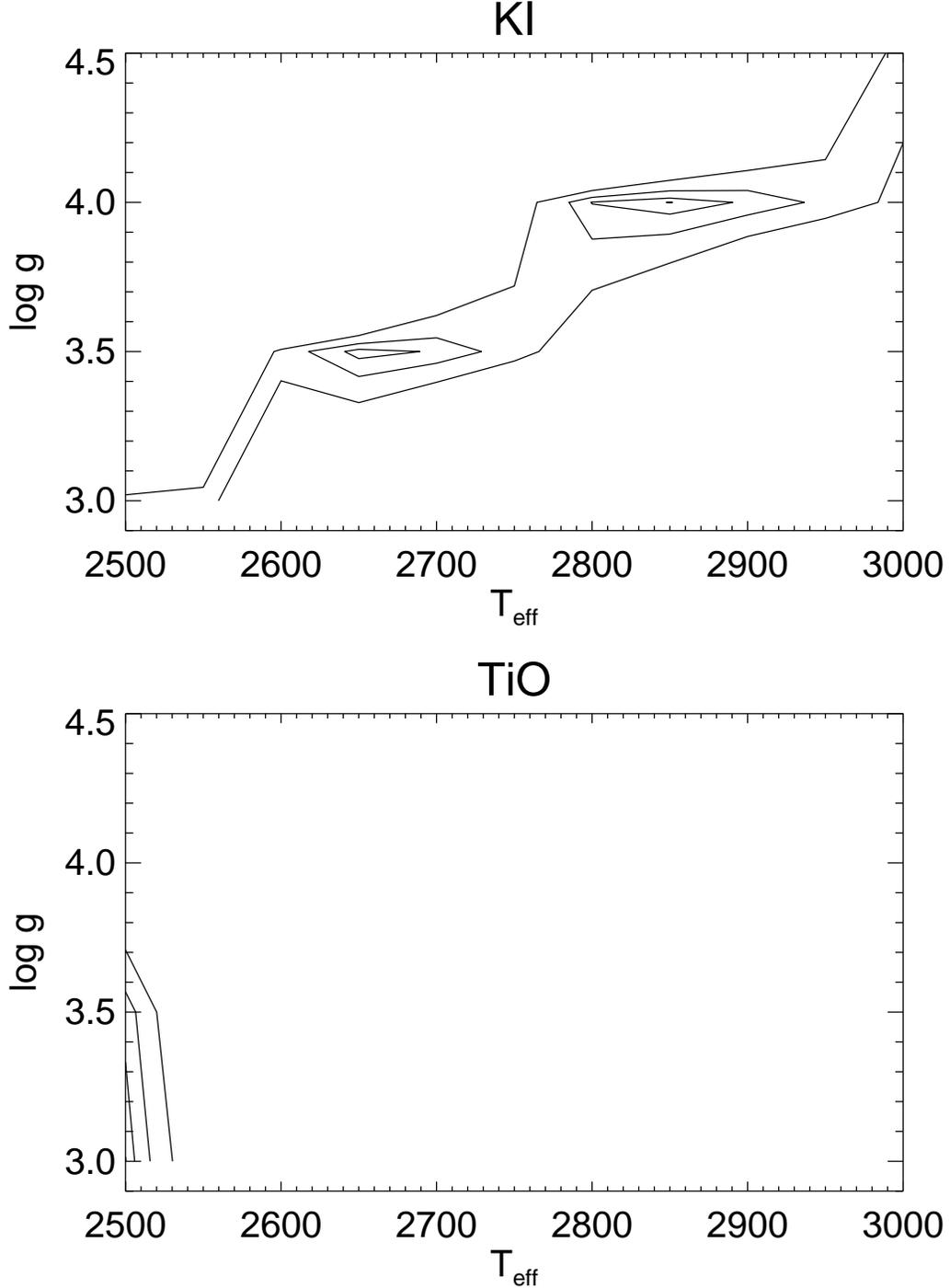}
\caption{\label{fig3} Determination of goodness-of-fit and formal
fit parameter uncertainties. Contours of constant $\chi^2 - \chi_{\rm
min}^2 =$ 2.3,6.2,11.8, representing 1,2,3-$\sigma$ joint confidence
intervals in the $T_{\rm eff}$--$\log g$ parameter plane. {\it Top:}
Joint confidence intervals for fitting of \ion{K}{1}. The absolute
minimum $\chi^2$ best fit is for $T_{\rm eff} = 2850$ K and $\log g =
4.0$, however a second equally good fit within 1-$\sigma$ confidence
occurs at $T_{\rm eff} = 2650$ K and $\log g = 3.5$. {\it Bottom:}
Joint confidence intervals for fitting of TiO. The contours demonstrate
that for TiO the best fitting model spectra are relatively insensitive
to $\log g$ but highly sensitive to $T_{\rm eff}$; a best-fit $T_{\rm
eff} = 2500 \pm 50$ is strongly preferred at high confidence. }
\end{figure}

\begin{figure}
\epsscale{0.75}
\plotone{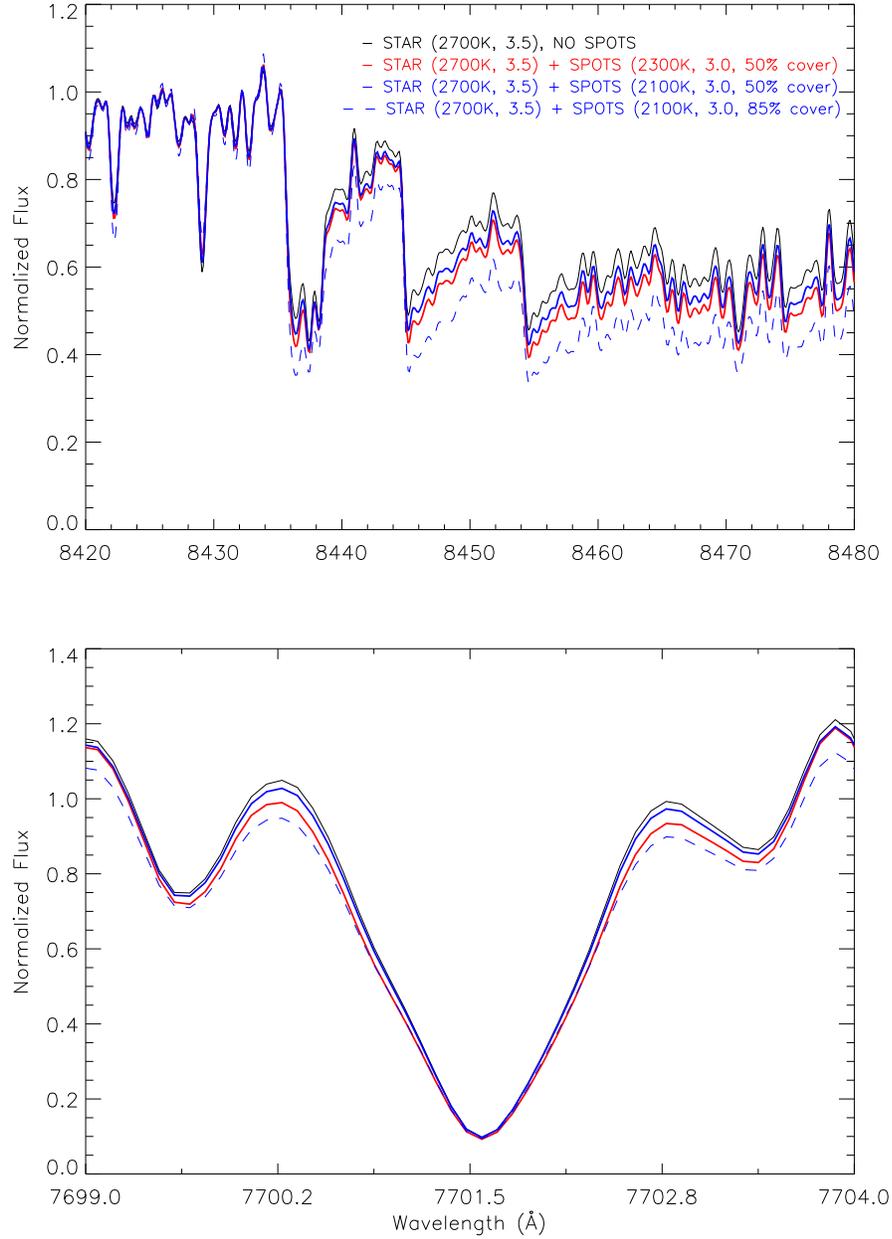}
\caption{\label{fig4} Trends in TiO ({\it top panel}) and KI
({\it bottom panel}) for spotted photospheres with varying spot
temperature and areal coverage.  For all models in both panels,
the unspotted photosphere has \teff\ = 2700K and \logg\ = 3.5,
while the spots have an effective gravity of 3.0.  In both panels,
the {\it black solid line} shows the unspotted model; the {\it red
solid line} shows the model with [T$_{SPOT}$, coverage] = [2300K,
50\%]; the {\it blue solid line} shows the model with [T$_{SPOT}$,
coverage] = [2100K, 50\%]; and the {\it blue dashed line} shows the
model with [T$_{SPOT}$, coverage] = [2100K, 85\%].  We see that the
[2300K, 50\%] spot model is deeper in TiO and broader in Ki than the
unspotted case, but the 2100K spot model with the same areal coverage
of 50\% has reversed in strength, and is closer to the unspotted
case than the 2300K model.  Increasing the areal coverage of the
2100K case to 85\%, however, makes it significantly deeper in TiO
and broader in KI than both the unspotted and 2300K spotted models.
These changes illustrate the spot-related trends discussed in \S 6.3.
This figure is shown in color in the electronic version only.}
\end{figure}

\begin{figure}
\epsscale{0.75}
\plotone{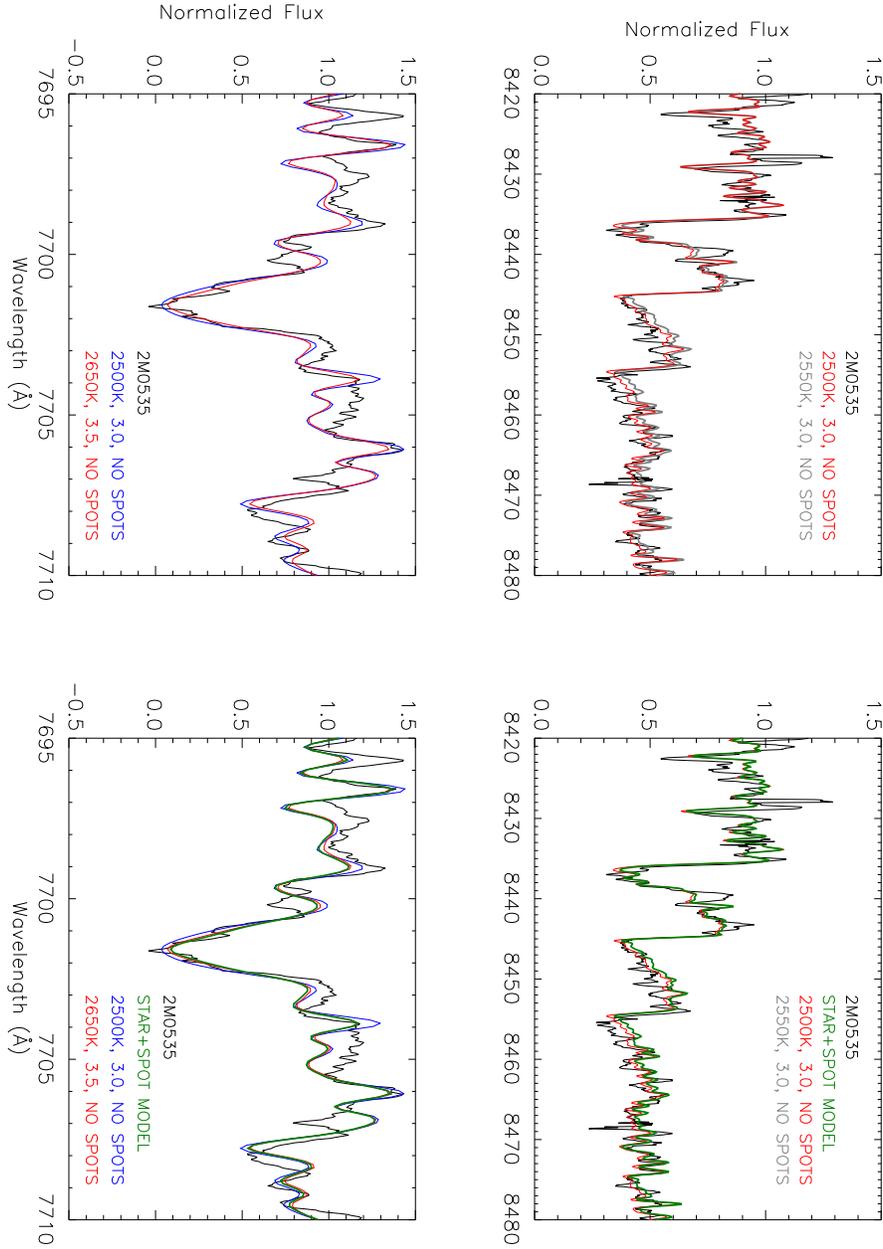}
\caption{\label{fig5} Data for 2M0535A ({\it black}) compared to
unspotted models ({\it left panels}) and a star+spot model ({\it
right panels}).  The TiO comparisons are shown at the {\it top}
and the K I comparisons at the {\it bottom}.  The unspotted models
plotted are the same best-fit ones plotted in Figs. 1 and 2 (one
best-fit model for TiO (in {\it red}), two for K I (in {\it red} and
{\it blue})).  Additionally, we plot a 2550K model (in {\it grey})
for TiO, to illustrate our adopted error bar of 50K in the \teff\
derived from TiO.  The spotted model (in {\it green} in the right
panels) is a [2700K, \logg\ = 3.5] photosphere with [2300K, log $g$
= 3.0] spots covering 70\% of the surface.  The fits to the spotted
model are nearly indistinguishable from the fits to the unspotted
models (specifically, the spotted model is an excellent match to the
2550K unspotted model in TiO, i.e., within our adopted \teff\ error
bar for TiO, and to the [2650K, \logg\ = 3.5] unspotted model in KI),
implying that the spotted model is as good a description of the data
as the unspotted ones.  See \S6.3. 
This figure is shown in color in the electronic version only.}
\end{figure}

%\clearpage

%\begin{deluxetable}{lccc}
%\tablecaption{Broadband Flux Measurements of 2M0535-05\label{obs-table}}
%\tablehead{
%\colhead{Passband} & 
%\colhead{$\lambda$} & 
%\colhead{$F_\lambda$} &\\
% & $\mu$m & mJy}

%\startdata
%$g'$   &   0.48   &   0.017$\pm$0.002 \\
%MIPS   &    24    &   $<$1.69\tablenotemark{1} \\
%\enddata
%\tablenotetext{1}{3$\sigma$ upper limit.}
%\end{deluxetable}

\end{document}